\documentclass[pra,twocolumn,showpacs,floatfix,a4paper]{revtex4}
\usepackage{graphicx}
\usepackage{amsmath}
\usepackage{amssymb}
\usepackage{amsfonts}
\usepackage{color}

\DeclareMathOperator{\sinc}{sinc}
\DeclareMathOperator{\rect}{rect}

\begin{document}

\title{Tunable linear and quadratic optomechanical coupling for a tilted membrane within an optical cavity: theory and experiment}

\author{M. Karuza}
\affiliation{School of Science and Technology, Physics Division, University of Camerino, via Madonna delle Carceri, 9, I-62032 Camerino (MC), Italy, and INFN, Sezione di Perugia, Italy}
\author{M. Galassi}
\affiliation{School of Science and Technology, Physics Division, University of Camerino, via Madonna delle Carceri, 9, I-62032 Camerino (MC), Italy, and INFN, Sezione di Perugia, Italy}
\author{C. Biancofiore}
\affiliation{School of Science and Technology, Physics Division, University of Camerino, via Madonna delle Carceri, 9, I-62032 Camerino (MC), Italy, and INFN, Sezione di Perugia, Italy}
\author{C. Molinelli}
\affiliation{School of Science and Technology, Physics Division, University of Camerino, via Madonna delle Carceri, 9, I-62032 Camerino (MC), Italy, and INFN, Sezione di Perugia, Italy}
\author{R. Natali}
\affiliation{School of Science and Technology, Physics Division, University of Camerino, via Madonna delle Carceri, 9, I-62032 Camerino (MC), Italy, and INFN, Sezione di Perugia, Italy}
\author{P. Tombesi}
\affiliation{School of Science and Technology, Physics Division, University of Camerino, via Madonna delle Carceri, 9, I-62032 Camerino (MC), Italy, and INFN, Sezione di Perugia, Italy}
\author{G. Di Giuseppe}
\affiliation{School of Science and Technology, Physics Division, University of Camerino, via Madonna delle Carceri, 9, I-62032 Camerino (MC), Italy, and INFN, Sezione di Perugia, Italy}
\author{D. Vitali}
\affiliation{School of Science and Technology, Physics Division, University of Camerino, via Madonna delle Carceri, 9, I-62032 Camerino (MC), Italy, and INFN, Sezione di Perugia, Italy}

\begin{abstract}
We present an experimental study of an optomechanical system formed by a vibrating thin semi-transparent membrane within a high-finesse optical cavity. We show that the coupling between the optical cavity modes and the vibrational modes of the membrane can be tuned by varying the membrane position and orientation. In particular we demonstrate a large quadratic dispersive optomechanical coupling in correspondence with avoided crossings between optical cavity modes weakly coupled by scattering at the membrane surface. The experimental results are well explained by a first order perturbation treatment of the cavity eigenmodes.
\end{abstract}


\maketitle

\section{Introduction}

The study of cavity optomechanics has recently sparkled the interest of a broad scientific community due to its different applications, ranging from sensing of masses, forces and displacements at the ultimate quantum limits \cite{Schwab2005,Kippenberg2007}, to the realization of quantum interfaces for quantum information networks \cite{Mancini2003,Pirandola2004,Hammerer2009,Rabl2010}, up to tests of the validity of quantum mechanics at macroscopic level \cite{Marshall2003,Romero-Isart2011}. A large variety of devices has been recently proposed and tested in which a driven cavity mode interacts with a mechanical resonator due to the fact that the cavity mode frequency $\omega$ depends upon the effective position of the mechanical element $z$ \cite{Gigan2006,Arcizet2006,Kippenberg2007,Li2008,Thompson2008,Gavartin2011}. In most cases one has a linear dependence between $\omega$ and $z$, so that the optical field exerts an homogeneous force on the resonator, associated with either radiation pressure or the gradient dipole force, which has been proposed and used for cooling the resonator motion \cite{Marquardt2007,Wilson-Rae2007,Genes2008,Groblacher2009,Riviere2010,Teufel2011a,Chan2011}. In this case the phase shift of the output light is proportional to the mechanical displacement and one can implement high-sensitive readout of forces and displacements \cite{Teufel2009a,Anetsberger2010}. In the so-called membrane-in-the-middle scheme of Ref.~\cite{Thompson2008} the position dependence of the cavity mode frequency is caused by a semi-transparent thin membrane placed within the cavity. In such a case, when the membrane is placed at a node or at an antinode of the cavity field, $\omega(z)$ is \emph{quadratic} in the position of the mechanical element, and one has a dispersive interaction, which allows for new nonlinear optomechanical phenomena \cite{Nunnenkamp2010}, and for the quantum non-demolition detection of the vibrational quanta of the membrane motion \cite{Thompson2008,Clerk2010}.

Here we present an experimental study of the membrane-in-the-middle scheme. We show how the optomechanical coupling can be tuned by varying the position and the orientation of the membrane with respect to the cavity axis. In particular, we show that by appropriately tilting the membrane one can couple the various Hermite-Gauss modes of the cavity and induce avoided crossings between the frequency shifts of the cavity modes (see also Ref.~\cite{Sankey2010}). When the splitting at an avoided crossing is very small, the corresponding value of the second-order derivative of the frequency with respect to the position $\partial^2 \omega(z)/\partial z^2$ becomes very large and one achieves a significative quadratic dispersive optomechanical interaction. Here we experimentally achieve $\partial^2 \omega(z)/\partial z^2= 2\pi \times 4.46 $ MHz/nm$^2$, which is comparable to the value achieved in Ref.~\cite{Sankey2010}. The paper will also explain the experimental results by means of a first order perturbation theory able to illustrate how the tilted membrane determines the new cavity eigenmodes and their frequency shifts.

The outline of the paper is as follows. In Sec.~II we illustrate the main aspects of the membrane-in-the-middle scheme, while in Sec.~III we present our experimental setup. In Sec.~IV we present the experimental results on the tunability of the optomechanical coupling and the realization of a strong quadratic coupling, while Sec.~V is for concluding remarks. In the Appendix we provide details on the perturbation theory able to explain qualitatively and quantitatively the modification of the cavity modes caused by the membrane.

\section{The membrane-in-the-middle scheme}

The membrane-in-the-middle system is formed by a Fabry-Perot optical cavity with a thin semi-transparent membrane inside. The empty cavity supports an infinite set of optical modes, conveniently described by the Hermite-Gauss modes~\cite{Siegman1986}. The thin membrane is a dielectric slab of thickness $L_d$ and complex index of refraction $n_M=n_R+\mathrm{i}n_I$. When it is placed within the cavity at a position $z_0$ measured along the cavity axis, the mode functions and their frequency change in a way which is dependent upon the position and orientation of the membrane with respect to the cavity~\cite{Thompson2008,Wilson2009,Sankey2010,Biancofiore2011}.

Considering the membrane motion means assuming that its mean position along the cavity axis oscillates in time, $z_0\rightarrow z_0+z(x,y,t)$, where $z(x,y,t)$ gives the membrane transverse deformation field, and is given by a superposition of the vibrational normal modes. One typically assumes the high stress regime of a taut membrane, in which bending effects are negligible and the classical wave equation well describes these normal modes~\cite{Wilson-Rae2011,Biancofiore2011}. The membrane vibrational modes are coupled to the optical cavity modes by radiation pressure, and therefore one has in general a multimode bosonic system in which many mechanical and optical modes interact in a nonlinear way. However, one can often adopt a simplified description based on a \emph{single} cavity mode interacting with a \emph{single} mechanical mode~\cite{Thompson2008,Biancofiore2011,Jayich2008}. This is possible when: i) the driving laser mainly populates a single cavity mode (with annihilation operator $\hat{a}$), and scattering into other modes is negligible~\cite{Law1995}); ii) the detection bandwidth is chosen so that it includes only a single, isolated, mechanical resonance with frequency $\Omega_m$ (described by dimensionless position $\hat{q}$ and momentum $\hat{p}$ operators, such that $[\hat{q},\hat{p}]= i$). By explicitly including cavity driving by a laser with frequency $\omega_L$ and input power ${\mathcal P}$, one ends up with the following cavity optomechanical Hamiltonian,
\begin{equation}
H=\frac{\hbar \Omega _{m}}{2}(\hat{p}^{2}+\hat{q}^{2})+\hbar \omega(\hat{q})\hat{a}^{\dagger }\hat{a}
+i\hbar E(a^{\dagger }e^{i\omega_L t}-ae^{-i\omega_L t}),  \label{eq:Ham-optomech}
\end{equation}
where $E=\sqrt{2\mathcal{P}\kappa_0 /\hbar \omega _{L}}$, with $\kappa_0$ the cavity mode bandwidth in the absence of the membrane.
In Eq.~(\ref{eq:Ham-optomech}) we have included the radiation pressure interaction within the cavity mode energy term, by introducing a position-dependent cavity frequency $\omega(\hat{q})$, which can be written as
\begin{equation}\label{eq:pos-dep-freq-gen}
\omega(\hat{q})=\omega_0+\delta\omega\left[z_0(\hat{q}), \alpha_x, \alpha_y\right],
\end{equation}
where $\omega_0$ is the cavity mode frequency in the absence of the membrane, and $\delta\omega\left[z_0(\hat{q}), \alpha_x, \alpha_y\right]$ is the frequency shift caused by the insertion of the membrane. This shift depends upon the membrane position along the cavity axis $z_0(\hat{q})$, which in turn depends upon the coordinate $\hat{q}$ because $z_0(\hat{q}) = z_0+x_0 \Theta \hat{q} $, where $z_0$ is the membrane center-of-mass position along the cavity axis, $x_0=\sqrt{\hbar/m \Omega_m}$ is the spatial width of the mechanical zero point motion ($m$ is the effective mass of the mechanical mode), and $\Theta$ is the dimensionless overlap integral between the transverse mode functions of the selected mechanical and optical modes~\cite{Biancofiore2011}. The frequency shift also depends upon $\alpha_x$ and $\alpha_y$, the tilting angles around the $x$ and $y$ axis respectively. The $(x,y,z)$ axes form a left-handed cartesian frame with the origin at the center of the cavity, and $x$ and $y$ are otherwise arbitrary, due to the cylindrical symmetry of the optomechanical system around the cavity axis $z$.

When the thin membrane is perfectly aligned, $\alpha_x=\alpha_y=0$, and is placed very close to the cavity waist, the Hermite-Gauss modes still represent the cavity eigenmodes with a very good approximation, because their wavefronts fit well with the membrane. The frequency shift in this case is a simple periodic function of $z_0$, which is maximum at the antinodes and minimum at the nodes of the intracavity field, and mode degeneracy is not removed \cite{Jayich2008,Biancofiore2011}. When the membrane is appreciably shifted from the waist and/or is tilted, light scattering of the Hermite-Gauss modes at the membrane surface is no more negligible and the cavity eigenmodes are significantly modified~\cite{Sankey2010}. When the longitudinal shift from the waist and the tilting angles are sufficiently small, one can describe the situation by means of a degenerate first-order perturbation theory of the wave equation within the cavity. In this perturbation limit, the new eigenmodes are linear combinations of few Hermite-Gauss modes and the corresponding frequency shifts $\delta\omega\left[z_0(\hat{q}), \alpha_x, \alpha_y\right]$ can be correspondingly evaluated~\cite{Sankey2008}. In this paper we will study both experimentally and theoretically the behavior of these frequency shifts: we will see in particular that the cavity mode frequencies, and therefore the optomechanical coupling as well, can be fine-tuned and controlled \emph{in situ} by varying the position and the orientation of the membrane.

In fact, the optomechanical coupling between the selected cavity and membrane vibrational modes is provided by the first-order (and eventually higher-order) term in the expansion of $\delta\omega\left[z_0(\hat{q}), \alpha_x, \alpha_y\right]$ as a function of $\hat{q}$. At most membrane positions $z_0$ a first order expansion in $\hat{q}$ of Eq.~(\ref{eq:pos-dep-freq-gen}) provides an accurate description of the physics: in this case one has a standard radiation pressure optomechanical interaction with a single-photon optomechanical coupling strength given by
\begin{equation}
G_0= \left|\frac{\partial \omega(\hat{q})}{\partial \hat{q}}\right|=\left|\frac{\partial \omega}{\partial z_0}\right|x_0 \Theta,
\end{equation}
where $\partial \omega/\partial z_0$ can be directly measured experimentally, while $x_0$ and $\Theta$ depend upon the chosen optical and mechanical modes.

Instead, when the membrane center $z_0$ is placed exactly at a node or at an antinode of the cavity field, or at an avoided crossing between nearby frequencies, the first-order term in the expansion of $\omega(\hat{q})$ vanishes, and one has to consider the higher-order term, which is quadratic in $\hat{q}$. This latter term describes a dispersive interaction between the optical and the vibrational modes whose coupling rate is given by the second order derivative $\partial^2 \omega(\hat{q})/\partial \hat{q}^2$. This unique property of the membrane-in-the-middle scheme has been first discussed in \cite{Thompson2008,Jayich2008} and if such a quadratic coupling is sufficiently strong, could be exploited for a quantum non-demolition measurement of the vibrational energy \cite{Thompson2008,Clerk2010}.

\section{The experimental setup}\label{sec:setup}

Our membrane-in-the-middle setup is schematically described in Fig.~1. Laser light is produced
by a Nd:YAG laser (Innolight) with wavelength $\lambda = 1064$ nm. The light
passes through an optical isolator, and is sent to the optomechanical cavity via two steering mirrors. The cavity is formed by two
dielectric mirrors each with a radius of curvature $R_1=R_2=10$ cm (coated by Advanced Thin Films) and separated by a distance $L=9$ cm. The measured cavity finesse without the membrane is equal to $F_0\simeq 66000$.
The membrane is mounted on a piezo-motor driven optical mount that
controls the angular alignment of the membrane. The optical mount is in
turn mounted on xyz stack of piezo-motor driven linear stages that are
used for moving the membrane in space. The stages are vacuum compatible,
two of them are used for centering the membrane with respect to the
optical axis, while the remaining one positions the membrane along it.
Rotation around two axes perpendicular to the optical one provides us
with tip and tilt control.

\begin{figure}[ht]
\includegraphics[width=0.45\textwidth]{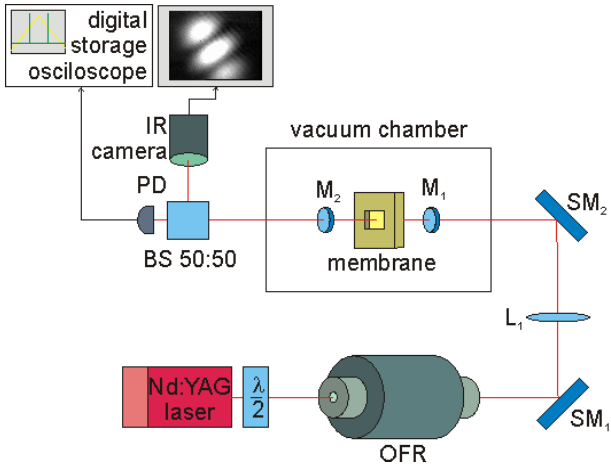}
\caption{Schematic description of the experimental setup.
 }
\label{fig:1}
\end{figure}

The membrane used in the experiment is a commercial, 1 mm $\times$ 1 mm ${\rm Si_3 N_4}$
stoichiometric x-ray window (Norcada), with nominal thickness $L_d =50$ nm, and index of refraction $n_R \simeq 2$. The membrane is supported by a 200 $\mu$m
thick Si frame, and it has been chosen due to its high mechanical quality factor and very low optical absorption ($n_I \sim 10^{-6}$ at $\lambda = 1064$ nm), as discussed in Ref.~\cite{Zwickl2008}. This parameter corresponds to an intensity reflection coefficient ${\cal R}=0.18$, which has been also experimentally verified.

The Fabry--Perot optical cavity is mounted inside a custom made vacuum
chamber which hosts also the membrane. The chamber is pumped down with a
turbo-molecular pump down to 10$^{-5}$ mbar. Once the base pressure is
reached, the pumping is switched off and the chamber is disconnected from
the pumping station, and the system is left in static vacuum. Usually
during the measurements the static vacuum inside the chamber reaches
values as high as 10$^{-2}$ mbar.

\section{Measurement of the cavity mode frequency shifts}

\begin{figure}[ht]
\includegraphics[width=0.49\textwidth]{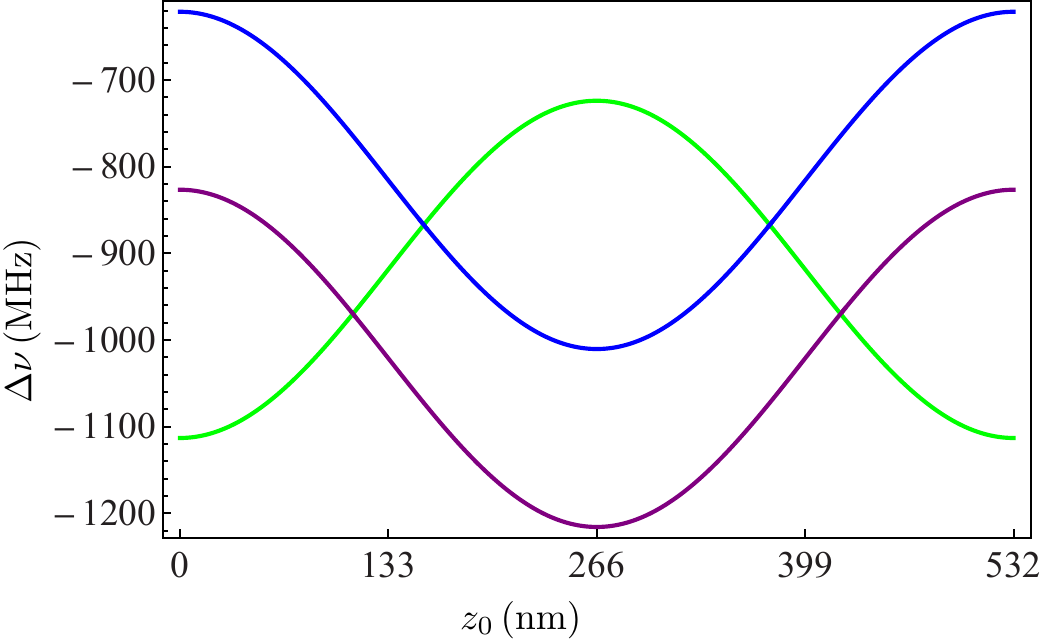}
\caption{Predicted frequency shift $\Delta \nu = \delta \omega/2\pi$ of the Hermite-Gauss modes close to the selected TEM$_{00,p}$ (blue curve, $p=$ longitudinal number) vs the membrane position along the cavity axis, for a perfectly aligned membrane around the waist. The green curve refers to the TEM$_{20,p-1}$, TEM$_{11,p-1}$, TEM$_{11,p-1}$ degenerate triplet, and the violet curve to the TEM$_{40,p-2}$, TEM$_{31,p-2}$, TEM$_{22,p-2}$, TEM$_{13,p-2}$, TEM$_{04,p-2}$ degenerate quintuplet. See Sec.~\protect\ref{sec:setup} for the other system parameters.
 }
\label{fig:2}
\end{figure}

We measure the optical cavity modes excited by the driving laser by monitoring the optical power transmitted through the cavity using the
photodiode shown in Fig.~1. We can also determine the transverse profile of the cavity mode
by imaging the transmitted beam with a video camera. The laser is tuned and mode-matched to one of the cavity's TEM$_{00}$ mode, and its frequency is then scanned over a prefixed range. In such a case, and if the input power is not too low, also higher order transverse TEM$_{mn}$ modes (with not too large $m$ and $n$), which are close in frequency to the driven TEM$_{00}$ mode, are populated and can be detected. For the system parameters given in Sec.~\ref{sec:setup}, the relevant cavity modes are shown in Fig.~2, which refers to a perfectly aligned membrane $\alpha_x=\alpha_y=0$, whose position is scanned for one half of the laser wavelength around the cavity waist.

\begin{figure}[ht]
\includegraphics[width=0.49\textwidth]{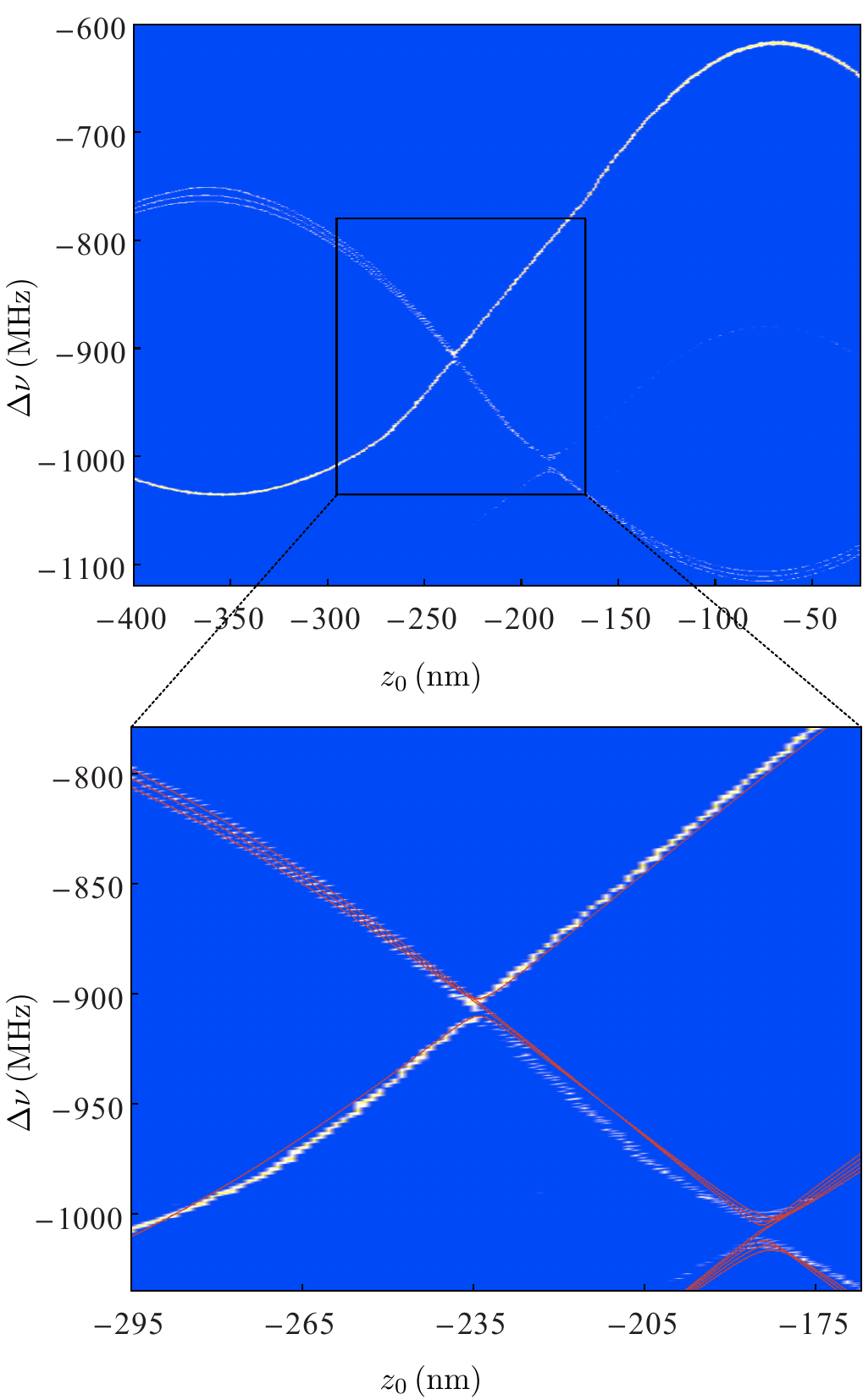}
\caption{Optical power transmitted through the cavity as a function of the laser frequency and the membrane position, in the case of membrane tilting angle $\alpha_x=-0.21$ mrad, $\alpha_y =0.15$ mrad, and shifted by $0.5$ mm from the waist. The crossings of the TEM$_{20,p-1}$ triplet with the TEM$_{00,p}$ mode and with the TEM$_{40,p-2}$ quintuplet are visible, together with the removed degeneracies and the presence of avoided crossings. Red curves in the zoomed part refer to the theoretical prediction of the perturbation theory developed in the Appendix.
 }
\label{fig:3}
\end{figure}

As discussed above, if the membrane is slightly misaligned, $\alpha_x,\alpha_y \neq 0$, the Hermite-Gauss TEM$_{mn}$ modes are no more cavity eigenmodes, and the frequency shifts are consequently modified. In particular: i) degeneracies are removed; ii) avoided crossings appear in correspondence of cavity modes which are coupled by the perturbation caused by the tilted membrane. This fact has been experimentally verified and it is shown in Fig.~3, where the optical power transmitted through the cavity as a function of the laser frequency and the membrane position is shown. Fig.~3 refers to a $0.5$ mm shift from the cavity waist and to tilting angles $\alpha_x=-0.21$ mrad, $\alpha_y =0.15$ mrad. The crossings of the TEM$_{20,p-1}$ triplet with the TEM$_{00,p}$ mode and with the TEM$_{40,p-2}$ quintuplet are visible, together with the removed degeneracies ($p$ denotes the longitudinal number of the driven TEM$_{00}$ mode). In the zoomed part avoided crossings are visible and their behavior is satisfactorily reproduced by the red curves, which correspond to the theoretical prediction of the perturbation theory developed in the appendix. In particular, the upper avoiding crossing concerns the TEM$_{00,p}$ and TEM$_{20,p-1}$ modes because the other two modes of the triplet, TEM$_{11,p-1}$ and TEM$_{02,p-1}$, are not coupled to the TEM$_{00,p}$ by the perturbation caused by the membrane. In the lower part of the figure one has instead three avoided crossings, associated to three pairs of modes which are coupled by the membrane perturbation, i.e., the $\left\{{\rm TEM}_{20,p-1},{\rm TEM}_{40,p-2}\right\}$, $\left\{{\rm TEM}_{11,p-1},{\rm TEM}_{31,p-2}\right\}$, and $\left\{{\rm TEM}_{02,p-1},{\rm TEM}_{04,p-2}\right\}$ pairs. In fact, the tilted membrane does not couple other pairs of triplet and quintuplet modes. This behavior is consistent with the fact that cylindrical symmetry around the cavity axis is satisfied at a very good level of approximation by our setup. We have checked this fact experimentally, by verifying that one gets essentially the same results if the values of $\alpha_x$ and $\alpha_y$ are exchanged. This means that in practice, for not too large misalignments, only a single tilting angle around an appropriate transverse axis, $\alpha_{eff}=\sqrt{\alpha_x^2+\alpha_y^2}$, is relevant.

Fig.~3 shows that, as expected, for most membrane position along the cavity axis, the frequency shift $\Delta \nu=\delta\omega/2\pi$ of the various optical modes has a linear dependence on $z_0$ and therefore on $\hat{q}$, corresponding to the usual radiation pressure optomechanical interaction. Fig.~3 also shows that the largest coupling strength is achieved for the TEM$_{00,p}$ mode for $z_0$ values halfway between a node and an antinode, and it is given by $|\partial \omega/\partial z_0|=2\pi \times 2.8 $ MHz/nm, which essentially coincides with the maximum achievable value $|\partial \omega/\partial z_0|_{max}=2\omega_0 \sqrt{{\cal R}}/L$ (see Ref.~\cite{Biancofiore2011}). The membrane we are currently investigating has a vibrational mode with effective mass $m\simeq 34$ ng and $\Omega_m\simeq 380$ kHz, corresponding to a zero-point motion width $x_0\simeq 1.11 \times 10^{-6}$ nm which, assuming an optimized overlap integral $\Theta \simeq 1$, implies a single-photon optomechanical coupling $G_0/2\pi \simeq 3.3$ Hz.

At membrane positions $z_0$ exactly corresponding to nodes, antinodes and avoided crossings, the linear term in $\hat{q}$ is zero and one has a purely quadratic dispersive optomechanical interaction,
\begin{equation}\label{eq:disp}
    H_{disp}=\hbar G_2 \hat{a}^{\dagger}\hat{a}\hat{q}^2, \;\;\;\;\;G_2=\frac{\partial^2 \omega}{\partial z_0^2}x_0^2.
\end{equation}
From Fig.~3 one gets that $|\partial^2 \omega/\partial z_0^2|\simeq 2 \pi \times 24$ kHz/nm$^2$ at nodes and antinodes of the TEM$_{00,p}$ mode, while one has a significant increase, by almost two orders of magnitude, at the avoided crossing between the TEM$_{00,p}$ mode and the triplet, where we measured $|\partial^2 \omega/\partial z_0^2|\simeq 2 \pi \times 1.0$ MHz/nm$^2$. This shows that avoided crossings allow to achieve significant values of the dispersive optomechanical coupling $G_2$ \cite{Sankey2010}, and suggests that by appropriately adjusting the membrane shift from the cavity waist and its tilting angle, one could ``engineer'' the spectrum of the optical cavity modes and generate avoided crossings with a very large value of the second order derivative $|\partial^2 \omega/\partial z_0^2|$. In fact, $|\partial^2 \omega/\partial z_0^2|$ is inversely proportional to the frequency splitting at the avoided crossing point, and therefore one should look for avoided crossings with a very small splitting, that is, between two cavity modes which are very weakly coupled by light scattering at the membrane surface.

\begin{figure}[ht]
\includegraphics[width=0.45\textwidth]{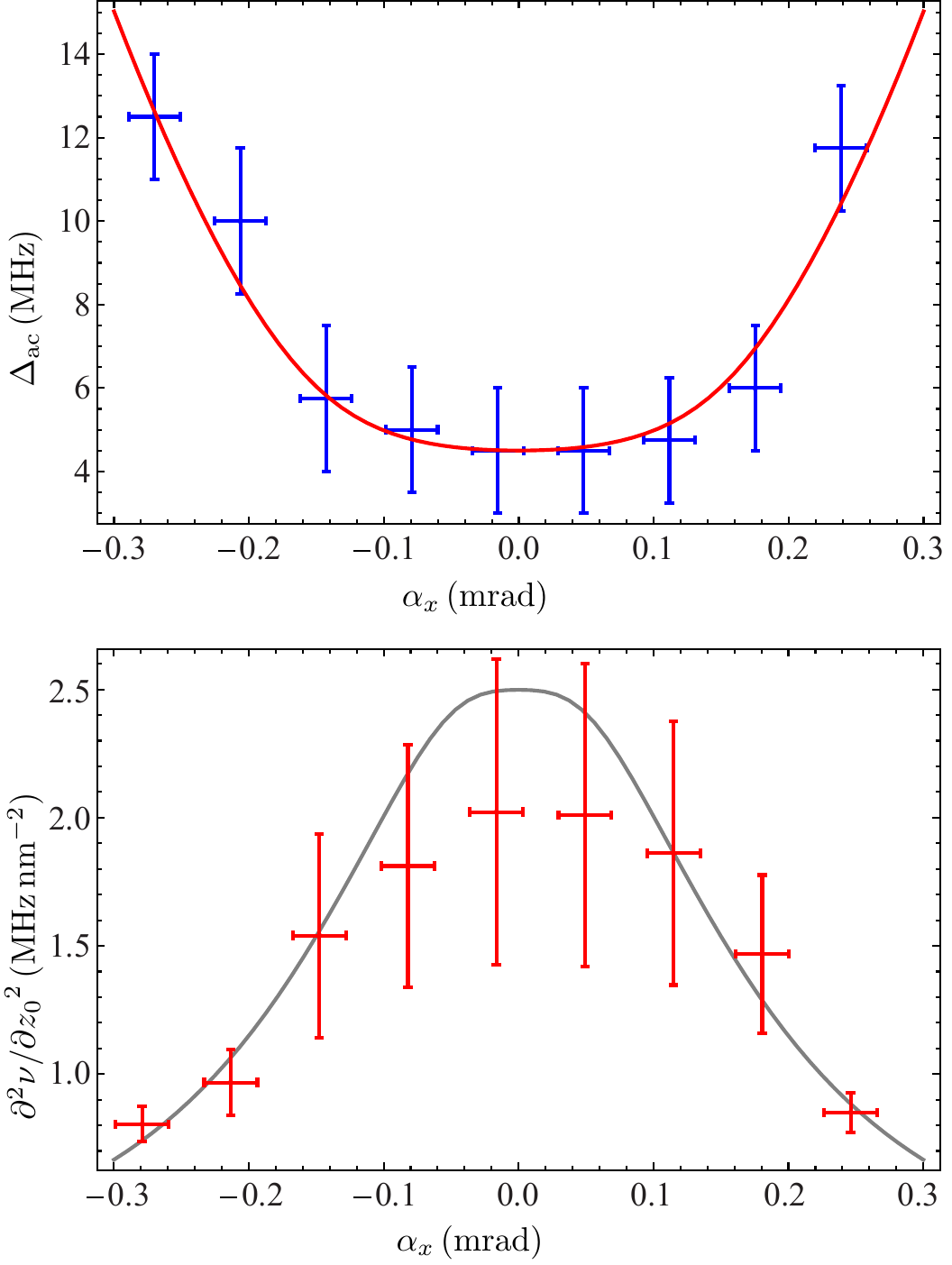}
\caption{Frequency splitting between the TEM$_{00,p}$ and the TEM$_{20,p-1}$ modes at the avoided crossing, $\Delta_{ac}$ (a), and second order derivative of the cavity mode frequency (b) vs the tilting angle $\alpha_x$. The other parameters are the same as in Fig.~3. In both figures the full curves refer to the theoretical prediction of the perturbation theory developed in the Appendix.
 }
\label{fig:4}
\end{figure}

We have first experimentally investigated how the frequency splitting and the associated second order derivative at the avoided crossing varies as a function of the tilting angle $\alpha_x$, while keeping all the other parameters fixed. The coupling between the TEM$_{00,p}$ and the TEM$_{20,p-1}$ modes decreases for decreasing $|\alpha_x|$, and therefore one expects to find a minimum of the splitting $\Delta_{ac}$, and a maximum of the second order derivative around $\alpha_x =0$. This behavior is confirmed by the experimental data shown in Fig.~4, which refer to the avoided crossing between the TEM$_{00,p}$ and the TEM$_{20,p-1}$ modes, $\Delta_{ac}$, and to the same parameters of Fig.~3, except that we have varied $\alpha_x$ around $\alpha_x = 0$. In particular we see that around $\alpha_x \simeq 0$ we achieve $|\partial^2 \omega/\partial z_0^2|\simeq 2 \pi \times 2.0$ MHz/nm$^2$. 

We have then looked for larger values of the quadratic optomechanical coupling in different parameter regions. We have found an interesting configuration by combining the effect of increasing the shift of the membrane position from the cavity waist at the center with that of membrane tilting. When the membrane is appreciably shifted from the waist, the curved wavefronts of the cavity modes do not fit anymore with the flat membrane surface, and one has an additional perturbation which, differently from membrane tilting, does not break the cylindrical symmetry around the cavity axis. The combined effects of the two perturbations may lead to very small couplings and therefore to avoided crossings with large second order derivative. Such a situation is shown in Fig.~5, which corresponds to tilting angles $\alpha_x=0.77$ mrad, $\alpha_y \simeq 0$, and to a membrane shifted by $1.2$ mm from the waist. The dots refer to the measured values of the frequency shifts of the appreciably populated modes. The prediction of the perturbation theory developed in the appendix (given by the full lines in Fig.~5) well reproduces the frequency shifts provided that the quintuplet modes are included, even if they are never significantly populated.

\begin{figure}[ht]
\includegraphics[width=0.48\textwidth]{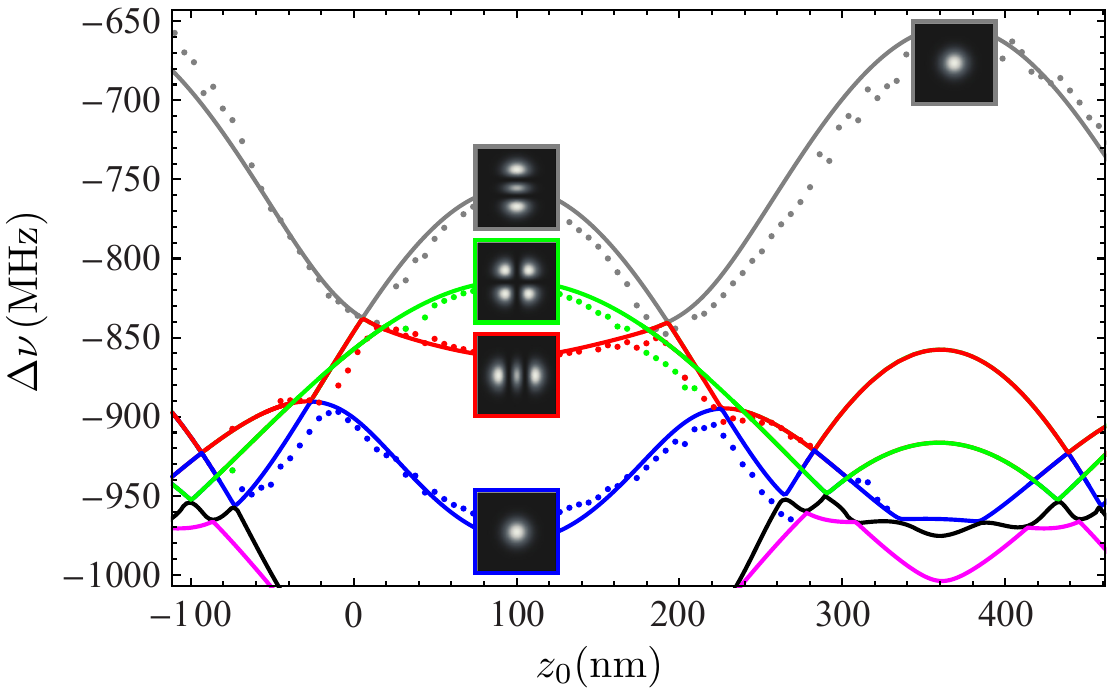}
\caption{The dots refer to the measured central peak frequency shifts $\Delta \nu=\delta\omega/2\pi$ of the appreciably populated cavity modes vs the membrane position along the cavity axis. The four different colors, grey, red, blue and green, correspond to the populated modes and can be associated with the TEM$_{00,p}$ mode and the triplet modes, TEM$_{20,p-1}$, TEM$_{11,p-1}$, TEM$_{02,p-1}$, only far from crossing points (see the corresponding transverse mode images). The full lines corresponds to theoretical prediction of the perturbation theory developed in the appendix for the corresponding membrane configuration, which is $\alpha_x=0.77$ mrad, $\alpha_y \simeq 0$, and shifted by $1.2$ mm from the waist. The quintuplet modes are not appreciably populated in this case, but they need to be taken into account in the perturbation theory in order to reproduce satisfactorily the frequency shifts and some of them are in fact shown in the plot. The avoided crossing of interest, with a very small splitting, is the one between the grey and red curve around $z_0 = 0$.
 }
\label{fig:5}
\end{figure}

\begin{figure}[ht]
\includegraphics[width=0.48\textwidth]{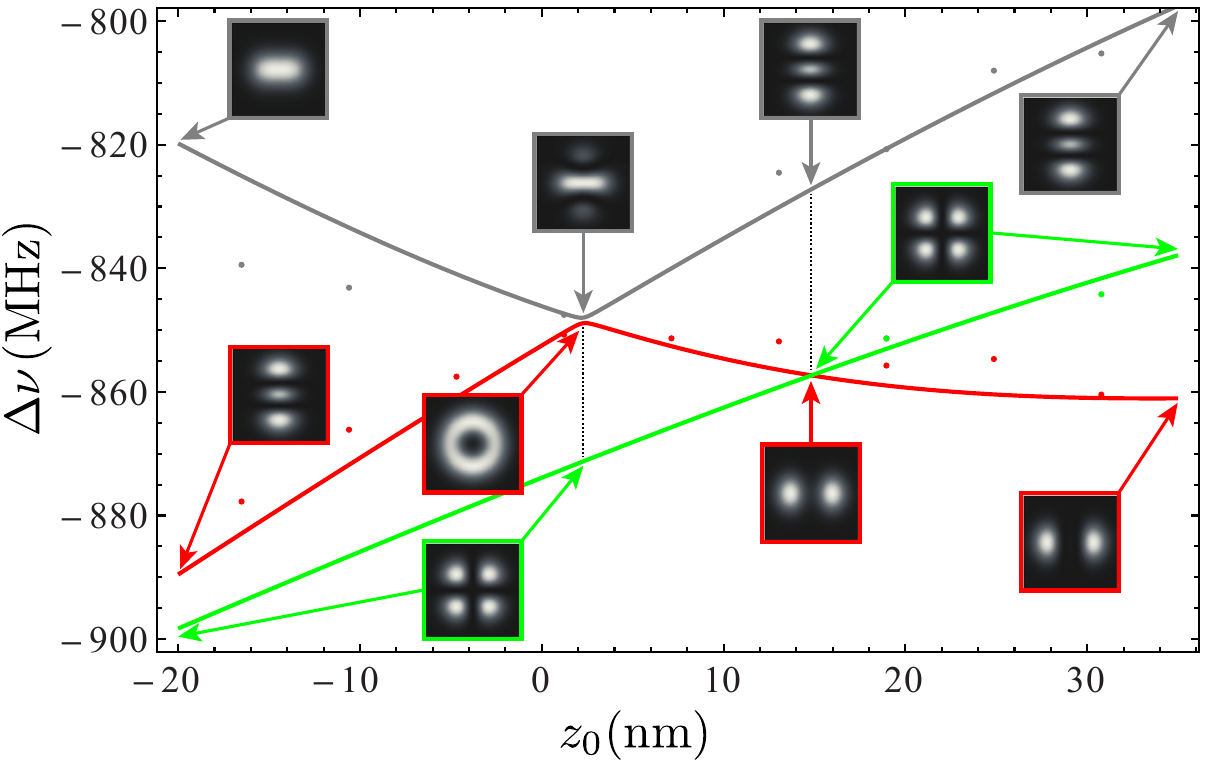}
\caption{Zoom of Fig.~5 around the avoided crossing point in the vicinity of the position $z_0=0$. The measured frequency shifts (dots), the theoretical predictions (full lines) are shown, together with the image of the transverse pattern of the new cavity eigenmodes in various membrane positions.
 }
\label{fig:6}
\end{figure}

The interesting point in Fig.~5 is the avoided crossing between the grey and the red curve, around $z_0 = 0$, which is characterized by a very small splitting. The zoom of this part is shown in Fig.~6 where, together with the measured and predicted frequency shifts, we show the images of the transverse pattern of the modes at various membrane positions. The splitting at the avoided crossing is of the order of $1$ MHz and, more important, one has at this point $|\partial^2 \omega/\partial z_0^2| = 2\pi \times 4.46$ MHz/nm$^2$, which is significantly larger than the values at nodes and antinodes, and is comparable to the value measured in Ref.~\cite{Sankey2010}. In this avoided crossing point, the splitting is very small because the combined effect of tilting and shift from the waist causes a small but nonzero coupling between two orthogonal linear combinations of triplet and the TEM$_{00}$ modes. These two cavity eigenmodes have the transverse pattern shown in Fig.~6, and with a good approximation they can be written as $|\phi\rangle_{red}= \left[|TEM_{20,p-1}\rangle+|TEM_{02,p-1}\rangle-|TEM_{00,p}\rangle \right]/\sqrt{3}$, and $|\phi\rangle_{gray}= \left[2|TEM_{02,p-1}\rangle-|TEM_{20,p-1}\rangle+|TEM_{00,p}\rangle \right]/\sqrt{6}$.

\section{Conclusions}

We have studied both theoretically and experimentally a membrane-in-the-middle setup formed by a high-finesse Fabry-Perot cavity with a thin semi-transparent SiN membrane inside. We have seen that the position and orientation of the membrane within the cavity allows to fine-tune the frequencies of the cavity modes, and through it, the optomechanical coupling of these modes with the vibrational modes of the membrane. In most membrane positions the frequency shifts of the cavity mode depends linearly upon the membrane deformation and therefore one has the traditional radiation pressure coupling between optical and mechanical modes. However at the nodes and antinodes of the cavity field the linear term vanishes and one has a dispersive interaction, which is quadratic in the position operator of the mechanical mode. We have demonstrated that such a quadratic coupling can be enhanced by two orders of magnitude in correspondence of avoided crossings with small frequency splitting. We have demonstrated a quadratic coupling term comparable to that achieved in \cite{Sankey2010}, associated with an avoided crossing between two new cavity eigenmodes which are linear combinations of the TEM$_{00,p}$, TEM$_{20,p-1}$, and TEM$_{02,p-1}$ modes.

\section{Acknowledgments}

This work has been supported by the European Commission (FP-7 FET-Open project MINOS).

\begin{appendix}

\section{First order degenerate perturbation theory}  \label{bigappen}

To describe the intracavity electromagnetic field in the presence of a slightly tilted membrane we adopt a degenerate first-order perturbation theory similar to that discussed in Ref.~\cite{Sankey2008} and start from the Gauss-Hermite modes of the empty cavity as zero-th order solution. We consider a symmetric cavity formed by two identical spherical mirrors with radius of curvature $R$, separated by a distance $L$, in the coordinate system with the $z$ axis along the cavity axis and centered at the cavity center. The thin membrane is a dielectric slab of thickness $L_d$ and complex index of refraction $n_M=n_R+\mathrm{i}n_I$.

The time-independent wave equation of the empty intracavity electromagnetic field is
\begin{equation}\label{eq:we}
 (\nabla^2+k_j^2)\tilde{\phi}_j(r)=0,
\end{equation}
where we have defined $\tilde{\phi}_j(r)= \sqrt{2}\,{\rm Re}[\phi_j(r)]$ for future convenience. $j$ is a collective index corresponding to the triplet of natural numbers $(l_j,m_j,n_j)$, which determines the \mbox{$j$-th} Gauss-Hermite mode \cite{Siegman1986}.

Defining $\phi_j^{(s)}=\begin{cases}\phi_j,&\text{if }s=1\\ \phi_j^*,&\text{if }s=-1\end{cases}$, we have
\begin{equation}
 \phi_j^{(s)}(r)=\rho_j(r)\mathrm{e}^{-s\mathrm{i}\theta_j(r)},
\end{equation}
with
\begin{eqnarray}
  \rho_j(r)&=&\frac{\mathrm{H}_{m_j}\bigl[\frac{\sqrt{2}x}{w_j(z)}\bigr]\mathrm{H}_{n_j}\bigl[\frac{\sqrt{2}y}{w_j(z)}\bigr]\exp\bigl[-\frac{x^2+y^2}{w_j(z)^2}\bigr]}{\sqrt{\pi\,2^{m_j+n_j-1}m_j!n_j!L}},\\
  \theta_j(r)&=&k_j z-(m_j+n_j+1)\arctan\Bigl(\frac{z}{z_R}\Bigr)+\\ \nonumber &&+\,\frac{x^2+y^2}{w_j(z)^2}\frac{z}{z_R}+(l_j-1)\pi/2.
\end{eqnarray}
$\mathrm{H}_m$ is the \mbox{$m$-th} Hermite polynomial, $w_j(z)=w_j^{(0)}\sqrt{1+(z/z_R)^2}$ gives the \mbox{$j$-th} Gaussian beam transversal shape along $z$ axis, $w_j^{(0)}=\sqrt{2 z_R/k_j}$ is the \mbox{$j$-th} Gaussian beam radius at the cavity waist, with
$z_R=(L/2)\sqrt{(1+g)/(1-g)}$ the Rayleigh range of the optical cavity, and $k_j=(\pi/L)\bigl[l_j+(m_j+n_j+1)\arccos(g)/\pi\bigr]$ is the norm of the \mbox{$j$-th} wave vector \cite{Siegman1986}.
The set of functions $\bigl\{\tilde{\phi}_j^{(s)}\bigr\}$ forms an approximate complete set of orthonormal functions for the space region within the Fabry-Perot cavity $\mathcal{G}$, also satisfying null boundary conditions at the surface of the end cavity mirrors, that is,
 \begin{eqnarray}\label{eq:ortho}
  \int_\mathcal{G}\mathrm{d}^3 r\,\tilde{\phi}_i(r)\tilde{\phi}_j(r)&\approx&\delta_{ij}=\delta_{l_i l_j}\delta_{m_i m_j}\delta_{n_i n_j},\\
  \tilde{\phi}_j(r)\big\arrowvert_{r\in\mathsf{M_\pm}}&\approx& 0.
 \end{eqnarray}
The insertion of the membrane in a tilted and shifted position modifies the time-independent electromagnetic field wave equation in this way
\begin{equation}\label{eq:mod_we}
 \bigl\{\nabla^2+k^2\bigl[1+V(r)\bigr]\bigr\}\tilde{\psi}(r)=0,
\end{equation}
where $V(r)=\bigl(n_M^2-1\bigr)\rect\bigl\{\bigl[z-z_c(x,y)\bigr]/t_{\alpha_{xy}}\bigr\}$, where $\rect(\zeta)$ is defined as $\rect(\zeta)=0$ if $\lvert\zeta\rvert>1/2$, and $\rect(\zeta)=1$ if $\lvert\zeta\rvert<1/2$; moreover $t_{\alpha_{xy}}$ is the \textit{corrected thickness} defined as $t_{\alpha_{xy}}= L_d/\cos\bigl[\bigl(\alpha_x^2+\alpha_y^2\bigr)^{1/2}\bigr]$.  $z_c(x,y)=z_0+\alpha_x x+\alpha_y y$ gives, approximately, the $z$-axis projection of a generic point, of transversal coordinate $(x,y)$,  belonging to the (near orthogonal to $z$ axis) central plane of symmetry of the membrane. The $V$ function gives therefore the perturbation caused by the presence of the membrane within the cavity.

We write the solution $\tilde\psi$ of the equation \eqref{eq:mod_we} as a linear combination of the orthonormal basis functions $\bigl\{\tilde{\phi}_j\bigr\}$
\begin{equation}\label{eq:mod_psi}
 \tilde\psi=\sum_j c_j\tilde{\phi}_j.
\end{equation}
Substituting \eqref{eq:mod_psi} in \eqref{eq:mod_we}, using \eqref{eq:we}, multiplying by $\tilde{\phi}_i$, integrating over $\mathcal{G}$ and using \eqref{eq:ortho}, we obtain the equations for each $i$,
\begin{equation}\label{eq:syst}
 c_i\bigl(1-k_i^2/k^2\bigr)+\sum_j c_j V_{ij}=0,
\end{equation}
with $V_{ij}= \int_\mathcal{G}\mathrm{d}^3r\,\tilde{\phi}_i(r)V(r)\tilde{\phi}_j(r)$, the matrix elements of the $V$ function in the chosen orthonormal basis (see subsection \ref{ap:matrix_el} for an approximated explicit calculation of $V_{ij}$).

In the simpler case of two degenerate cavity modes with frequencies $\omega_1$ and $\omega_2$ and with degeneracies $n_1$ and $n_2$ respectively, the system of equations \eqref{eq:syst} is a finite system with $n= n_1+n_2$ equations
\begin{equation}\label{eq:finite_syst}
 c_i\bigl(1-k_i^2/k^2\bigr)+\sum_{j=1}^n c_j V_{ij}=0,\quad i\in\bigl\{1,2,\dots,n\bigr\}.
\end{equation}
We have $k_j^2/k_1^2=1$ for $j\in\{1,2,\dots,n_1\}$, and we define $\eta= k_2^2/k_1^2=k_j^2/k_1^2 $ for $j\in\{n_1+1,n_1+2,\dots,n\}$, where typically  $\eta\approx 1$. Defining $\lambda^{-1}= k^2/k_1^2$ and $W_{jj} = 1+V_{jj}$ for $j\in\{1,2,\dots,n\}$, we can write \eqref{eq:finite_syst} in matrix form as
\begin{widetext}
 \begin{equation}\label{eq:syst_matrix}
  \begin{pmatrix}
  W_{1,1}-\lambda & \dots & V_{1,n_1} & V_{1,n_1+1} & \dots & V_{1,n} \\
  \vdots & \ddots & \vdots & \vdots & & \vdots \\
  V_{1,n_1} & \dots & W_{n_1,n_1}-\lambda & V_{n_1,n_1+1} & \dots & V_{n_1,n} \\
  V_{1,n_1+1} & \dots & V_{n_1,n_1+1} & W_{n_1+1,n_1+1}-\eta\lambda & \dots & V_{n_1+1,n} \\
  \vdots & & \vdots & \vdots & \ddots & \vdots \\
  V_{1,n} & \dots & V_{n_1,n} & V_{n_1+1,n} & \dots & W_{n,n}-\eta\lambda \\
 \end{pmatrix}
 \begin{pmatrix}
  c_1 \\
  \vdots \\
  c_{n_1} \\
  c_{n_1+1} \\
  \vdots \\
  c_n
 \end{pmatrix}=
 \begin{pmatrix}
  0 \\
  \vdots \\
  0 \\
  0 \\
  \vdots \\
  0
 \end{pmatrix},
 \end{equation}
which is a sort of an eigenvalue problem with respect to $\lambda$. Solving \eqref{eq:syst_matrix} gives $\lambda$ as a function of the membrane position specified by $(z_0,\alpha_x,\alpha_y)$. In particular we obtain the frequency shifts with respect to the frequency of the driving laser $\omega_0$,
\begin{equation}
 \delta\omega(z_0,\alpha_x,\alpha_y)=c k_1 [\lambda(z_0,\alpha_x,\alpha_y)]^{-1/2}-\omega_0.
\end{equation}
In the case of stoichiometric $\mathrm{Si}_3\mathrm{N}_4$ membrane one has a negligible optical absorption, that is $n_R\approx 2$ and $n_I \simeq 10^{-6}$. Defining the extinction coefficient $\kappa= n_I/n_R$, one has $\kappa\lesssim 10^{-5}\ll1$, and one can exploit the first order approximation
\begin{equation}
 \lambda[n_M(\kappa)]\approx\lambda[n_M(0)]+\bigl\{\partial_{n_M}\lambda[n_M(\kappa)]\bigr\}n_M'(\kappa)\bigr\rvert_{\kappa=0}\kappa=\lambda(n_R)+2\mathrm{i}n_I\lambda'(n_R).
\end{equation}
\end{widetext}
Using this approximation, we can calculate the $\lambda$'s as the real roots of the determinant of the matrix of coefficients in \eqref{eq:syst_matrix} with real entries (each element of the matrix with $n_M=n_R$).

\subsection{A singlet-triplet case study}

We apply now the general method exposed above to the particular case of a singlet mode coupled to a triplet of modes. That is, we assume, as it is relevant for the experimental system described in this paper, that the input laser drives a non-degenerate $\mathrm{TEM}_{00}$ mode with longitudinal index $l$, specified by  $\left\{(l,0,0)\right\}$, and associated with the non degenerate frequency $\omega_1=\left(\pi\mathrm{c}/L\right)\times$ $\bigl[l+$ $\arccos\left(g\right)/\pi\bigr]$. This mode is coupled to the triplet specified by the modes $\bigl\{(l-1,2,0),$ $(l-1,1,1),$ $(l-1,0,2)\bigr\}$, associated with the three-fold degenerate frequency $\omega_2=\left(\pi\mathrm{c}/L\right)\times$ $\bigl[\left(l-1\right)+$ $3\arccos\left(g\right)/\pi\bigr]$.

We have $\omega_1\approx$ $\omega_2\approx$ $\omega_L$ by hypothesis. As discussed in the text, due to the cylindrical symmetry with respect to the cavity axis, we can consider (without loss of generality) one of the two (small) angles, $\alpha_x$ and $\alpha_y$, to be zero (we choose $\alpha_y=0$). In particular, by defining $\zeta_0= z_0/z_R$, we consider two sub-cases: -- small arbitrary $\zeta_0$ and $\alpha_x$; -- small arbitrary $\zeta_0$ with $\alpha_x=0$.

\subsubsection{One-angle-tilted and shifted membrane}

If the membrane is generally not aligned (with $\alpha_y=0$) and not centered at the waist (small arbitrary $\zeta_0$ and $\alpha_x$), by applying \eqref{eq:syst_matrix}, we obtain the eigenvalue equation $\det[V_\eta(\lambda)]=0$ for $\lambda$, where
 \begin{equation}\label{eq:char1}
 V_\eta(\lambda)=\begin{pmatrix}
  W_{11}-\lambda & V_{12} & 0		& V_{14}	\\
  V_{12} & W_{22}-\eta\lambda & 0		& V_{24}		\\
  0      & 0      & W_{33}-\eta\lambda	& 0		\\
  V_{14} & V_{24}	  & 0		& W_{44}-\eta\lambda
 \end{pmatrix}.
\end{equation}
The solution of the eigenvalue equation gives: $\lambda_1=\bigl(1+V_{33}\bigr)/\eta$ and three others values $\bigl\{\lambda_1,$ $\lambda_2,$ $\lambda_3\bigr\}$ which are given by the roots of the cubic equation $\sum_{i=0}^3 a_i\lambda^{3-i}=0$, where:
\begin{eqnarray}
 a_0&=&\eta^2,\\
 a_1&=&-\eta\bigl(\eta W_{11}+W_{22}+W_{44}\bigr),\\
 a_2&=&-\eta V_{12}\bigl(V_{12}+V_{14}\bigr)-V_{24}^2+\eta W_{11} W_{22}+\nonumber\\
  &&+\,W_{44}\bigl(\eta W_{11}+W_{22}\bigr),\\
 a_3&=&V_{12}^2\bigl(W_{44}-V_{24}\bigr)+V_{14}V_{12}\bigl(W_{22}-V_{24}\bigr)+\nonumber\\
 &&+\,W_{11}\bigl(V_{24}^2-W_{22}W_{44}\bigr),\\
 W_{ii}
&=&1+V_{ii},\quad i\in\left\{1,\dots,4\right\}.
\end{eqnarray}

\subsubsection{Aligned and shifted membrane}

If the membrane is perfectly aligned ($\alpha_x=\alpha_y=0$) and not centered at the cavity waist (small arbitrary $\zeta_0$), the eigenvalue equation simplifies because only one off-diagonal matrix element is nonzero, corresponding to the (equal) coupling of the $\mathrm{TEM}_{00}$ mode with the $\mathrm{TEM}_{20}$ and $\mathrm{TEM}_{02}$ modes, i.e.,
\begin{equation}
 V_\eta(\lambda)=\begin{pmatrix}
  W_{11}-\lambda & V_{12} & 0		& V_{12}	\\
  V_{12} & W_{22}-\eta\lambda & 0		& 0		\\
  0      & 0      & W_{22}-\eta\lambda	& 0		\\
  V_{12} & 0	  & 0		& W_{22}-\eta\lambda
 \end{pmatrix}.
\end{equation}
The eigenvalues can be explicitly obtained and are given by
 \begin{eqnarray}
  \lambda_1 &=&\lambda_2\,\,=\,\,\frac{1+V_{22}}{\eta},\\
  \lambda_{3,4}&=&\frac{1}{2}\biggl\{\Bigl(1+V_{11}+\frac{1+V_{22}}{\eta}\Bigr)+\nonumber\\&&\,\pm\biggl[\frac{8V_{12}^2}{\eta}+\Bigl(1+V_{11}-\frac{1+V_{22}}{\eta}\Bigr)^2\biggr]^{1/2}\biggr\}.\quad\quad
 \end{eqnarray}
A further simplification is that in this case also the matrix elements $V_{ij}$ results less involved than the previous case with a nonzero tilting angle.

\begin{widetext}

\subsection{Approximate evaluation of the matrix elements}\label{ap:matrix_el}

It is possible to derive an approximated explicit expression for the matrix elements $V_{ij}$. We can write:
\begin{eqnarray}
&& V_{ij}(\zeta_0,\alpha_x,\alpha_y)=(n_M^2-1)\int_\mathcal{G}\mathrm{d}^3r\,{\rm Re}\bigl[\sqrt{2}\phi_i(r)\bigr]\zeta(r){\rm Re}\bigl[\sqrt{2}\phi_j(r)\bigr]\\
&&=(n_M^2-1){\rm Re}\Bigl[\int_\mathcal{G}\mathrm{d}^3r\, \phi_i(r)\zeta(r) \phi_j(r)+\int_\mathcal{G}\mathrm{d}^3r\, \phi_i(r) \zeta(r) \phi_j^*(r)\Bigr]=(n_M^2-1){\rm Re} \bigl[I^{(s=1)}_{ij}(\zeta_0,\alpha_x,\alpha_y)+I^{(s=-1)}_{ij}(\zeta_0,\alpha_x,\alpha_y)\bigr],\nonumber
\end{eqnarray}
where we have defined
\begin{equation}\label{eq:Isij}
 I^{(s)}_{ij}(\zeta_0,\alpha_x,\alpha_y)=\int_\mathcal{G}\mathrm{d}^3r\,\phi_i^{(s=1)}(r)\zeta(r)\phi_j^{(s)}(r).
\end{equation}
The problem is now reduced to the evaluation of the integral \eqref{eq:Isij} and of its real part in particular. Using the fact that $\zeta_0$ is small one gets (see also \cite{Sankey2008})
\begin{eqnarray}\label{eq:Isij2}
 I^{(s)}_{ij}(\zeta_0,\alpha_x,\alpha_y)&\approx&\bigl(\pi^2 4^{m_{ij}^{+}+n_{ij}^{+}}m_i!m_j!n_i!n_j!\bigr)^{-1/2}(t_{\alpha_{xy}}/L)\sinc\bigl( K^{(s)}_{ij}t_{\alpha_{xy}}/z_{R}\bigr)\mathrm{e}^{-K^{(s)2}_{ij}\bigl(\alpha^2_{xij}+\alpha^2_{yij}\bigr)}\mathrm{e}^{-\mathrm{i}\bigl(2 K^{(s)}_{ij}\zeta_0+\pi\ell^{(s)}_{ij}\bigr)}\nonumber\\
 &&\times \Bigl\{\Gamma^{(s)}_{ij00}-\mathrm{i}\Delta K^{(s)}_{ij}\Bigl[\bigl(\Gamma^{(s)}_{ij02}+\Gamma^{(s)}_{ij20}\bigr)\zeta_0+\bigl(\Gamma^{(s)}_{ij12}+\Gamma^{(s)}_{ij30}\bigr)
 \alpha_{xij}+\bigl(\Gamma^{(s)}_{ij03}+\Gamma^{(s)}_{ij21}\bigr)\alpha_{yij}\Bigr]\Bigr\},
\end{eqnarray}
where, by introducing $\delta\tilde{k}_{ij}=(k_{i}-k_{j})/(k_{i}+k_{j})$, $\kappa^{\pm}_{ij}=(k_{i}\pm k_{j})/(2 z_{R})$, we have used the notations $m^{\pm}_{ij}=(m_i\pm m_j)/2$, $n^{\pm}_{ij}=(n_i\pm n_j)/2$, $K^{(s)}_{ij}=(1+s)\bigl[\kappa^{+ }_{ij}-\bigl(m^{+}_{ij}+n^{+}_{ij}+1\bigr)\bigr]/2+(1-s)\bigl[\kappa^{-}_{ij}-\bigl( m^{-}_{ij}+n^{-}_{ij}\bigr)\bigr]/2$,
$\Delta K^{(s)}_{ij}=\left[1+s+(1-s)\delta\tilde{k}_{ij}\right]/2$,
$\alpha_{xij}=\alpha_x/\sqrt{\kappa^{+}_{ij}}$, $\alpha_{yij}=\alpha_y/\sqrt{\kappa^{+}_{ij}}$,
$\ell^{(s)}_{ij}=\left[\ell_i-1+s(\ell_j-1) \right]/2$, and also defined
\begin{equation}
\Gamma^{(s)}_{ijqp}(\alpha_x,\alpha_y) = J(m_i,m_j,K^{(s)}_{ij}\alpha_{xij},q,\delta\tilde{k}_{ij})J(n_i,n_j,K^{(s)}_{ij}\alpha_{yij},p,\delta\tilde{k}_{ij}),\label{eq:Gammasijqp}
\end{equation}
where
\begin{equation}
J(m,n,c,q,\delta)=\int_{-\infty}^{+\infty}\mathrm{d}x\,\mathrm{e}^{-x^2}(x-\mathrm{i}c)^q\,\mathrm{H}_m
\bigl[\sqrt{1+\delta}(x-\mathrm{i}c)\bigr]\mathrm{H}_n\bigl[\sqrt{1-\delta}(x-\mathrm{i}c)\bigr]= J^R(m,n,c,q,\delta)-\mathrm{i}J^I(m,n,c,q,\delta).
\end{equation}
This integral can be explicitly evaluated using the properties of the Hermite polynomials, and one gets
\begin{eqnarray}
  J^R(m,n,c,q,\delta)&=&\partial_{t_1}^m\partial_{t_2}^n\bigl\{g_\delta\bigl[\cos(2f_\delta c)R_e-\sin(2f_\delta c)R_o\bigr]\bigr\}_{t_1=t_2=0},\\
  J^I(m,n,c,q,\delta)&=&\partial_{t_1}^m\partial_{t_2}^n\bigl\{g_\delta\bigl[\sin(2f_\delta c)R_e+\cos(2f_\delta c)R_o\bigr]\bigr\}_{t_1=t_2=0},
 \end{eqnarray}
where $ f_\delta(t_1,t_2)= t_1\sqrt{1+\delta}+t_2\sqrt{1-\delta}$, $g_\delta(t_1,t_2)=\exp\left\{-t_1^2-t_2^2+f_\delta^2\right\}$, and
\begin{eqnarray}  R_e(q,f_\delta,c)&=&\sum_{k=0}^{\lfloor q/2\rfloor}(-1)^k \binom{q}{2k}S_{q-2k}(f_\delta)c^{2k},\\
  R_o(q,f_\delta,c)&=&\sum_{k=0}^{\lfloor (q-1)/2\rfloor}(-1)^k \binom{q}{2k+1}S_{q-(2k+1)}(f_\delta)c^{2k+1},
  \end{eqnarray}
with  $S_k(f_\delta)=\sum_{j=0}^{\lfloor k/2\rfloor}\binom{k}{2j}\Gamma(j+1/2)f_\delta^{k-2j}$. Using these results, one finally gets the following expression for the relevant integral of Eq.~(\ref{eq:Isij2}),
 \begin{eqnarray}
 && {\rm Re}[I_{ij}^{(s)}(\zeta_0,\alpha_x,\alpha_y)]=x_1\cos x_2\Bigl\{{\rm Re}\left[\Gamma_{ij00}^{(s)}\right]+\Delta K_{ij}^{(s)}\Bigl[{\rm Im}\left[\Gamma_{ij02}^{(s)}+\Gamma_{ij20}^{(s)}\right]\zeta_0+{\rm Im}\left[\Gamma_{ij12}^{(s)}+\Gamma_{ij30}^{(s)}\right]\alpha_{xij}\nonumber \\
 &&+{\rm Im}\left[\Gamma_{ij21}^{(s)}+\Gamma_{ij03}^{(s)}\right]\alpha_{yij}\Bigr]\Bigr\}+x_1\sin x_2\Bigl\{{\rm Im}\left[\Gamma_{ij00}^{(s)}\right]-\Delta K_{ij}^{(s)}\Bigl[{\rm Re}\left[\Gamma_{ij02}^{(s)}+\Gamma_{ij20}^{(s)}\right]\zeta_0 \nonumber \\
 &&+{\rm Re}\left[\Gamma_{ij12}^{(s)}+\Gamma_{ij30}^{(s)}\right]\alpha_{xij}+{\rm Re}\left[\Gamma_{ij21}^{(s)}+\Gamma_{ij03}^{(s)}\right]\alpha_{yij}\Bigr]\Bigr\}
 \end{eqnarray}
where we have defined
\begin{eqnarray}\label{eq:ReIsijComponents}
  x_1&= & \bigl(\pi^2 4^{m_{ij}^{+}+n_{ij}^{+}}m_i!m_j!n_i!n_j!\bigr)^{-1/2}(t_{\alpha_{xy}}/L)\sinc\bigl( K^{(s)}_{ij}t_{\alpha_{xy}}/z_{R}\bigr)\mathrm{e}^{-K^{(s)2}_{ij}\bigl(\alpha^2_{xij}+\alpha^2_{yij}\bigr)},\\
  x_2&=&2 K^{(s)}_{ij}\zeta_0+\pi\ell^{(s)}_{ij}.
\end{eqnarray}

\end{widetext}

\end{appendix}

\bibliography{optomechanics-membrane-exp}

\end{document}